\documentclass[showpacs,aps,graphicx,amsmath]{revtex4}

\usepackage{graphicx}
\usepackage{amsmath}
\usepackage{amssymb}

\usepackage{multirow}

\begin{document}

\title{ Fast universal quantum gates on microwave photons with all-resonance operations in circuit QED\footnote{Published in Sci. Rep. \textbf{5}, 9274 (2015)} }
\author{Ming Hua, Ming-Jie Tao, and Fu-Guo Deng\footnote{Correspondence and
requests for materials should be addressed to F. -G. Deng
(fgdeng@bnu.edu.cn).}}
\address{Department of Physics, Applied Optics Beijing Area Major Laboratory, Beijing normal University, Beijing 100875, China}
\date{\today }

\begin{abstract}

Stark shift  on a superconducting qubit in circuit quantum
electrodynamics (QED) has been used to construct universal quantum
entangling gates on superconducting resonators in previous works. It
is a second-order coupling effect between the resonator and the
qubit in the dispersive regime, which leads to a slow
state-selective rotation on the qubit. Here, we present two
proposals to construct the fast universal quantum gates on
superconducting resonators in a microwave-photon quantum processor
composed of multiple superconducting resonators coupled to a
superconducting transmon qutrit, that is,  the controlled-phase
(c-phase) gate on two microwave-photon resonators and the
controlled-controlled phase (cc-phase) gates on  three resonators,
resorting to   quantum resonance operations, without any drive
field.  Compared with previous works,  our universal quantum gates
have the higher fidelities and shorter operation times in theory.
The numerical simulation shows that the fidelity of our c-phase gate
is 99.57\% within about 38.1 ns and that of our cc-phase gate is
99.25\% within about  73.3 ns.
\end{abstract}

\maketitle






Quantum computation and quantum information processing have attached
much attention \cite{book} in recent years. A quantum computer can
factor an $n$-bit integer exponentially faster than the best known
classical algorithms and it can run the famous quantum search
algorithm, sometimes known as Grover's algorithm, which enables this
search method to be speed up substantially, requiring only
O($\sqrt{N}$) operations, faster than the classical one which
requires O($N$) operations \cite{book}. Universal quantum gates are
the key elements in a universal quantum computer, especially  the
controlled-phase (c-phase) gate or its equivalent gate -- the
controlled-not (CNOT) gate. C-phase gates (or CNOT gates) assisted
by single-qubit rotations can construct a universal quantum
computing.  Compared to the synthesis with universal two-qubit
entangling gates and single-qubit  gates, the direct implementation
of a universal three-qubit quantum gate [controlled-controlled-phase
(cc-phase) or controlled-controlled-not (Toffoli) gate] is more
economic and simpler as it requires at least six CNOT gates
\cite{Toffoli} to synthesize a  Toffoli gate which is equivalent to
a cc-phase gate.  By far, there are some interesting physical
systems used for the construction of universal quantum gates, such
as photons
\cite{photon1,photon2,photon3,photon4,photon5,photon6,photon7},
nuclear magnetic resonance
\cite{NMR,NMR2,longjcp,longprl,longpra04}, quantum dots
\cite{QD1,QD2,QD3,QD4,QD5,QD6,ZhangsPRA,QD7}, diamond
nitrogen-vacancy center \cite{NV1,NV2,NV3}, and cavity quantum
electrodynamics (QED) \cite{QED1,QED2}.

Circuit QED, composed of superconducting Josephson junctions (act as
the artificial atoms) and a superconducting resonator (acts as a
cavity and quantum bus)
\cite{SQ,Wallraff,AlexandreBlais,longglpra,Johansson,Ong,Rigetti},
is a promising implementation of cavity QED and it has the excellent
features of the good scalability and the long coherence time. It has
been used to realize the strong and even ultra-strong coupling
between a resonator and a superconducting qubit \cite{SQ,Forn}, and
complete some basic tasks of the quantum computation on the
superconducting qubits. For example, DiCarlo \emph{et al.}
\cite{DiCarlo} demonstrated the c-phase gate on two transmon qubits
assisted by circuit QED in 2009. In 2014, Chow \emph{et al.}
\cite{JerryMChow} experimentally implemented a strand of a scalable
fault-tolerant quantum computing fabric. In 2012, Fedorov \emph{et
al.} \cite{3q} implementated a Toffoli gate  and Reed \emph{et al.}
\cite{3q1} realized the three-qubit quantum error correction with
superconducting circuits. In 2014, Barends \emph{et al.}
\cite{Barends} realized the c-phase gate on every two adjacent Xmon
qubits with a high fidelity in a five-Xmon-qubit system assisted by
circuit QED. DiCarlo \emph{et al.} \cite{DiCarlo1} prepared and
measured the three-qubit entanglement in circuit QED in 2010, and
Steffen \emph{et al.} \cite{Steffen} realized the full deterministic
quantum teleportation with feed-forward in a chip-based
superconducting circuit architecture in 2013.

In a high-quality resonator, a microwave photon always has the
longer life time than that of a superconducting qubit
\cite{Devoret}, which makes the resonator  a  good candidate for
quantum information processing based on the basis of Fock states.
With a superconducting qubit coupled to a resonator, Hofheinz
\emph{et al.} \cite{Hofheinz} realized the generation of a Fock
state in 2008. In the same year, Wang \emph{et al.} \cite{HWang}
realized the measurement of the decay of Fock States. Hofheinz
\emph{et al.} \cite{Hofheinz2} demonstrated the synthesis of an
arbitrary superposition of Fock states in 2009. With two qubits
coupled to three resonators, Merkel and Wilhelm \cite{noon} proposed
a scheme for the generation of  the entangled NOON state on two
resonator qudits (with $d$ levels) in 2010. In 2011, Wang \emph{et
al.} \cite{Wang} demonstrated in experiment the generation of  the
entangled NOON state on two resonators. With a qubit coupled to two
resonators, Johnson \emph{et al.} \cite{Johnson} realized the single
microwave-photon non-demolition detection in 2010 and Strauch
\cite{Frederick1} exploited the all-resonant method to control the
quantum state of superconducting resonators and gave some theoretic
schemes for Fock state synthesis, qudit logic operations, and
synthesis of NOON states in 2012. With a qubit coupled to multiple
resonators, Yang \emph{et al.} proposed a theoretic scheme for the
generation of  the entangled Greenberger-Horne-Zeilinger state on
resonators based on the Fock states \cite{Siyuan} in 2012 and
entangled coherent states of four microwave resonators \cite{Han} in
2013.

Besides the entanglement generation for quantum information
processing, resonator qudits can also be used for quantum
computation, that is, universal quantum logic gates
\cite{FWStrauch,Wu,Hua}. In 2011, Strauch \cite{FWStrauch} gave an
interesting scheme to construct the quantum entangling gates on the
two resonator qudits based on the arbitrary Fock states, by using
the two-order coupling effect of the number-state-dependent
interaction between a superconducting qubit and a resonator in the
dispersive regime, discovered by Schuster \emph{et al.}
\cite{DISchuster} in 2007. The operation time of his c-phase gate on
two resonator qudits with the basis of the Fock states $|0\rangle_r$
and $|1\rangle_r$ is 150 ns. In a processor with one
two-energy-level charge qubit coupled to multiple resonators,  Wu
\emph{et al.} \cite{Wu} presented an effective scheme to construct
the  c-phase  gate on two resonators with  the
number-state-dependent interaction between the qubit and two
resonators in 2012. Its operation time is 125 ns. In a similar
processor, we gave a  scheme for the construction of the c-phase
gate  (cc-phase gate) on two (three) resonator qubits \cite{Hua}
(only working under the subspace of Fock states $\{|0\rangle_r,
|1\rangle_r \}$), by combining the number-state-dependent selective
rotation between a superconducting transmon qutrit (just the three
lowest energy levels are considered) and a resonator (two
resonators)  and the resonance operation on the rest resonator and
the qutrit in 2014.  The fidelity of our c-phase (cc-phase) gate can
reach 99.5\% (92.9\%) within 93 (124.6) ns in theory,  without
considering the decoherence and the dephasing rates of the qutrit
and the decay rate of the microwave-photon resonators.

In this paper, we  exploit  the all-resonance-based quantum
operations on a qutrit and  resonators to design two schemes for the
construction of the c-phase and the cc-phase gates on resonators in
a processor composed of multiple microwave-photon resonators coupled
to a transmon  qutrit, far different from the previous works for the
c-phase and cc-phase gates on resonators based on the second-order
couplings between the qubit and the resonators
\cite{FWStrauch,Wu,Hua}. Our simulation shows that the fidelity of
our c-phase gate on two microwave-photon resonators approaches
99.57\% within the operation time of about 38.1 ns and that of our
cc-phase gate on three resonators is 99.25\% within about 73.3 ns.
Our all-resonance-based universal quantum gates on
microwave-photon resonators without classical drive field are much faster than those in similar previous works \cite{FWStrauch,Wu,Hua}.\\

\begin{figure}[tpb]                 
\begin{center}
\includegraphics[width=7.0 cm,angle=0]{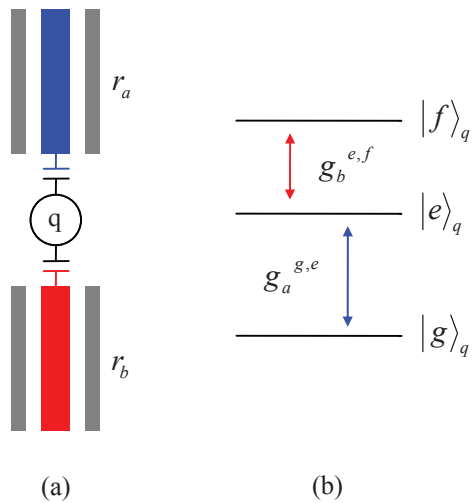} 
\end{center}
\caption{ (a) The schematic diagram  for our  c-phase gate on two
microwave-photon resontors in  circuit QED. It contains two
high-quality superconducting resonators ($r_a$ and $r_b$)
capacitively coupled to a superconducting quantum interferometer
device (SQUID) which acts as a superconducting transmon  qutrit
($q$), whose transition frequency can be tuned by the external flux.
(b) The schematic diagram for  the lowest three energy levels  of
the qutrit with small anharmonicity. $g_a^{g,e}$ is the coupling
strength between $r_a$ and the qutrit with the transition
$|g\rangle_q \leftrightarrow |e\rangle_q$. $g_b^{e,f}$ is the
coupling strength between $r_b$ and  the qutrit  with the transition
$|e\rangle_q \leftrightarrow |f\rangle_q$.} \label{fig1}
\end{figure}

\bigskip


{\large \textbf{Results}}

\textbf{ All-resonance-based c-phase gate on two resonator qubits.}
Let us consider the quantum  system  composed  of two high-quality
superconducting resonators coupled to a transmon qutrit  with the
three lowest  energy levels, i.e.,  the ground state $\left\vert
g\right\rangle _{q}$, the excited state $\left\vert e\right\rangle
_{q}$, and the second excited state $\left\vert f\right\rangle
_{q}$, shown in Fig. 1. In the interaction picture, the Hamiltonian
of the system is ($\hbar=1$):
\begin{eqnarray}             
\begin{split}
H_{2q} =  & g_{a}^{g,e}a\sigma _{g,e}^{+}e^{i\delta_{a}^{g,e}t}+g_{a}^{e,f}a\sigma _{e,f}^{+}e^{i\delta_{a}^{e,f}t}   \\
& +g_{b}^{g,e}b\sigma
_{g,e}^{+}e^{i\delta_{b}^{g,e}t}+g_{b}^{e,f}b\sigma
_{e,f}^{+}e^{i\delta_{b}^{e,f}t}+h.c.. \label{hamiltonian}
\end{split}
\end{eqnarray}
Here, $\sigma _{g,e}^{+}=\left\vert e\right\rangle _{q}\left\langle
g\right\vert $ and $\sigma _{e,f}^{+}=\left\vert f\right\rangle
_{q}\left\langle e\right\vert $ are the creation operators for the
transitions of the qutrit
 $\left\vert g\right\rangle _{q} \rightarrow\left\vert e\right\rangle
_{q}$ and  $\left\vert e\right\rangle _{q} \rightarrow\left\vert
f\right\rangle _{q}$, respectively. $a^{+}$ and $b^{+}$ are the
creation operators of the resonators $r_a$ and $r_b$ (labeled as $a$
and $b$ in subscripts), respectively.
$\delta_{a,(b)}^{g,e(e,f)}=\omega_{g,e(e,f)}-\omega_{a(b)}$ and
$\omega_{g,e(e,f)}=E_{e(f)}-E_{g(e)}$. $E_{i}$ is the energy for the
level $\left\vert i\right\rangle _{q}$ of the qutrit. $\omega _{a}$
and $\omega _{b}$ are the transition frequencies of the resonators
$a$ and $b$, respectively. $g_{a(b)}^{g,e}$ and $g_{a(b)}^{e,f}$ are
the coupling strengths between the resonator $r_a$ ($r_b$) and the
qutrit $q$ with  these two transitions.  Tuning the transition
frequencies of the transmon qutrit and the coupling strength between
the transmon qutrit and each resonator
\cite{Laloy,Harris,Srinivasan}, one can turn on and off the
interaction between the qutrit and each resonator effectively
\cite{Strauch}.

Let us suppose that the general initial state of the system is
\begin{eqnarray}               
\begin{split}
\vert \psi_{0}\rangle &= (\cos\theta_1|0\rangle_a+\sin\theta_1|1\rangle_a)\otimes(\cos\theta_2|0\rangle_b+\sin\theta_2|1\rangle_b)\otimes \vert g\rangle_{q}\\
&=(\alpha_1|0\rangle_a|0\rangle_b+\alpha_2|0\rangle_a|1\rangle_b+\alpha_3|1\rangle_a|0\rangle_b
+\alpha_4|1\rangle_a|1\rangle_b)\otimes \vert g\rangle_{q}.
\label{cphase0}
\end{split}
\end{eqnarray}
Here, $\alpha_1=\cos\theta_1\cos\theta_2$,
$\alpha_2=\cos\theta_1\sin\theta_2$,
$\alpha_3=\sin\theta_1\cos\theta_2$, and
$\alpha_4=\sin\theta_1\sin\theta_2$. The all-resonance-based c-phase
gate on two  microwave-photon resonators can be constructed with
three steps, shown in Fig. 2. We describe them in detail as follows.

\begin{figure}[tpb]              
\begin{center}
\includegraphics[width=9 cm,angle=0]{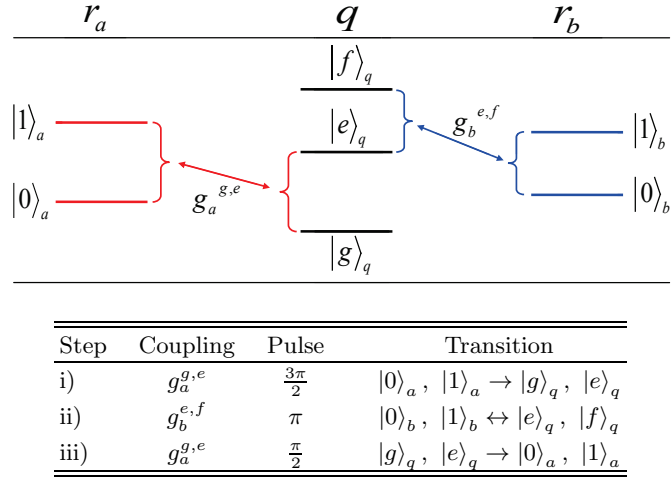} 
\end{center}
\begin{tabular}{lccccccccccc}
\hline\hline Step &  & & Coupling & &  & Pulse & & & & & Transition
\\ \hline i) & & & $g_{a}^{g,e}$ & &  & $\frac{3\pi }{2}$ & & & & &
$\left\vert 0\right\rangle _{a} \text{, }\left\vert 1\right\rangle
_{a}\rightarrow \left\vert g\right\rangle
_{q}\text{, }\left\vert e\right\rangle _{q}$ \\
ii) & & & $g_{b}^{e,f}$ & &  & $\pi $ & & & & & $\left\vert
0\right\rangle _{b}\text{, } \left\vert 1\right\rangle
_{b}\leftrightarrow \left\vert e\right\rangle _{q}
\text{, }\left\vert f\right\rangle _{q}$ \\
iii) & & & $g_{a}^{g,e}$ & &  & $\frac{\pi }{2}$ && & & &
$\left\vert g\right\rangle _{q} \text{, }\left\vert e\right\rangle
_{q}\rightarrow \left\vert 0\right\rangle _{a}\text{, }\left\vert
1\right\rangle _{a}$ \\ \hline\hline
\end{tabular}
\caption{The principle and the steps of our c-phase gate on $r_a$
and $r_b$ with  all-resonance operations. }\label{fig2}
\end{figure}

Step i), by tuning off the interaction between the transmon  qutrit
and $r_{b}$, and resonating $r_{a}$ and the two lowest energy levels
$\left\vert g\right\rangle _{q}$ and $\left\vert e\right\rangle
_{q}$ of the qutrit ($\omega_a = \omega_{g,e}$) with the operation
time of $t=\frac{3\pi}{2g_{a}^{g,e}}$, the state of the system can
be evolved into
\begin{eqnarray}           
\left\vert \psi_{1}\right\rangle \!=\! \left\vert
0\right\rangle_{a}\!\!\otimes\! (\alpha_1\left\vert
0\right\rangle_{b}\!\left\vert g\right\rangle_{q} \!\!+\!
\alpha_2\left\vert 1\right\rangle_{b}\!\left\vert g\right\rangle_{q}
\!\!+\! i\alpha_3\left\vert 0\right\rangle_{b}\!\left\vert
e\right\rangle_{q}
\!\!+\! i\alpha_4\left\vert 1\right\rangle_{b}\!\left\vert e\right\rangle_{q}).
\label{cphase1}
\end{eqnarray}

Step ii), by turning off the interaction between the qutrit and
$r_{a}$, and resonating $r_{b}$ and the  two energy levels
$\left\vert e\right\rangle _{q}$ and $\left\vert f\right\rangle
_{q}$ of the qutrit ($\omega_b = \omega_{e,f}$) with the operation
time of $t=\frac{\pi}{g_{b}^{e,f}}$, the state of the system can
evolve from $\left\vert \psi_{1}\right\rangle$ into
\begin{eqnarray}               
\left\vert \psi _{2}\right\rangle \!=\! \left\vert
0\right\rangle_{a} \!\!\otimes\! (\alpha_1\left\vert
0\right\rangle_{b}\!\left\vert g\right\rangle_{q} \!\!+\!\alpha_2
\left\vert 1\right\rangle_{b}\!\left\vert g\right\rangle_{q}   \!\!
+\! i\alpha_3\left\vert 0\right\rangle_{b}\!\left\vert
e\right\rangle_{q} \!\! - \! i\alpha_4\left\vert
1\right\rangle_{b}\!\left\vert e\right\rangle_{q}).
\label{cphase2}
\end{eqnarray}

Step iii), by turning off the interaction between the qutrit and
$r_b$,  and  resonating $r_{a}$ and the  energy levels $\left\vert
g\right\rangle _{q}$ and $\left\vert e\right\rangle _{q}$ of the
qutrit with the operation time of $t=\frac{\pi }{2g_{a}^{g,e}}$, one
can get the final state of the system as
\begin{eqnarray}               
\left\vert \psi_{f}\right\rangle \!=\!  \left\vert
g\right\rangle_{q}\!\otimes\!(\alpha_1\left\vert
0\right\rangle_{a}\!\left\vert 0\right\rangle _{b} \!+\!
\alpha_2\left\vert 0\right\rangle_{a}\!\left\vert 1\right\rangle_{b}
\!+\! \alpha_3\left\vert 1\right\rangle_{a}\!\left\vert
0\right\rangle_{b} \!-\! \alpha_4\left\vert
1\right\rangle_{a}\!\left\vert 1\right\rangle_{b}). \label{cphase3}
\end{eqnarray}
This is just the c-phase gate operation on the resonators $r_{a}$
and $r_{b}$, which indicates that a $\pi $ phase shift takes place
only if there is one microwave photon in each resonator.

If the operation time in step i) is taken to $t=\frac{\pi
}{2g_{a}^{g,e}}$, one can get the final state of the system as
\begin{eqnarray}                  
\left\vert \psi_{f}\right\rangle^{^{\prime }} \!\!=\! \left\vert
0\right\rangle_{q}\!\otimes\!\frac{1}{2}(\alpha_1\left\vert
0\right\rangle_{a}\!\left\vert
0\right\rangle_{b}\!+\!\alpha_2\left\vert
0\right\rangle_{a}\!\left\vert 1\right\rangle_{b}
\!-\!\alpha_3\left\vert 1\right\rangle_{a}\!\left\vert
0\right\rangle_{b} \!+\! \alpha_4\left\vert
1\right\rangle_{a}\!\left\vert 1\right\rangle_{b}). \label{cphase3'}
\end{eqnarray}
This is just the result of another c-phase gate operation on the
resonators $r_{a}$ and $r_{b}$, which indicates that the $\pi $
phase shift happens only if there is one microwave photon in $r_a$
and no microwave photon in $r_b$.

\begin{figure}[tpb!]               
\begin{center}
\includegraphics[width=16 cm,angle=0]{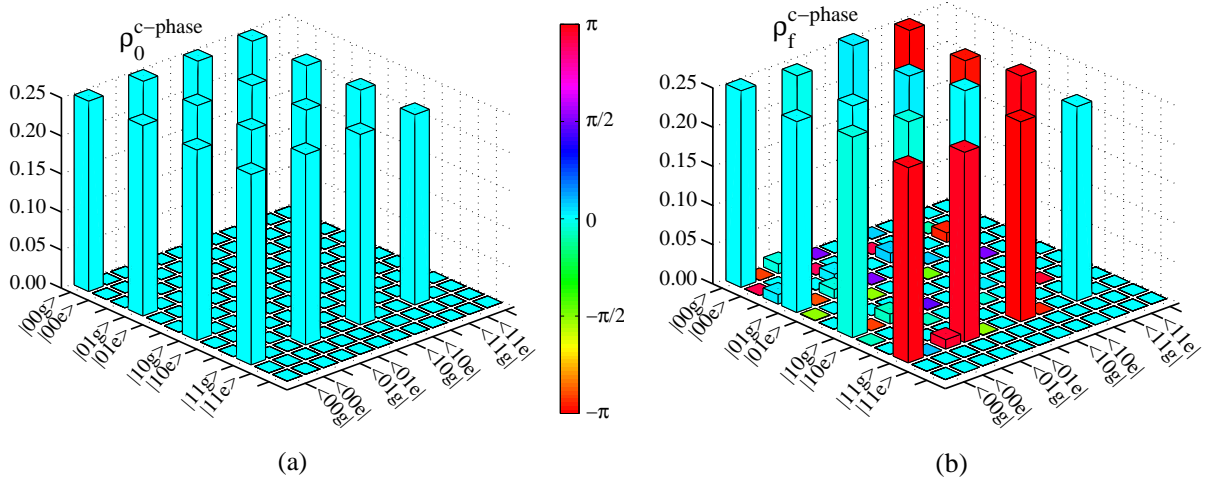}
\end{center}
\caption{(a) The density operator $\rho_0$ of the initial state
$\vert \psi_0\rangle$ of the quantum system composed of the two
resonator qudits and the superconducting qutrit for constructing our
c-phase gate. (b) The density operator $\rho_f^{c-phase}$ of the
final state $\left\vert \psi_{f}\right\rangle$ of the system.}
\label{fig3}
\end{figure}

To show the feasibility of the resonance processes for constructing
our c-phase gate, we numerically simulate its fidelity and operation
time with the feasible experiential parameters. The evolution of the
system composed of these two resonators and the transmon qutrit can
be described by the master equation \cite{SQ,Su}
\begin{eqnarray}              
\begin{split}
\frac{d\rho }{dt}  =&   -i[H_{2q},\rho ]+ \kappa_a D[a]\rho  +
\kappa_b D[b]\rho    +  \gamma_{g,e}D[\sigma_{g,e}^{-}]\rho
+\gamma_{e,f}D[\sigma_{e,f}^{-}]
\\ &   +  \gamma_{\phi,e}(\sigma_{ee}\rho\sigma_{ee}-\sigma_{ee}\rho/2-\rho\sigma_{ee}/2)
+
\gamma_{\phi,f}(\sigma_{ff}\rho\sigma_{ff}-\sigma_{ff}\rho/2-\rho\sigma_{ff}/2),
\label{masterequation}
\end{split}
\end{eqnarray}
where $D[L]\rho =(2L\rho L^{+}-L^{+}L\rho -\rho L^{+}L)/2$ with
$L=a$, $b$, $ \sigma _{g,e}^{-}$, $\sigma _{e,f}^{-}$.
$\sigma_{ee}=|e\rangle_q\langle e|$ and
$\sigma_{ff}=|f\rangle_q\langle f|$.  $\kappa_a$ ($\kappa_b$) is the
decay rate of the resonator  $r_a$ ($r_b$),  $\gamma_{g,e}$
($\gamma_{e,f}$) is the energy relaxation rate of the qutrit with
the transition $\vert e\rangle \rightarrow \vert g\rangle$ ($\vert
f\rangle \rightarrow \vert e\rangle$), and $\gamma_{\phi,e}$ and
$\gamma_{\phi,f}$ are the dephasing rates of the levels $\vert
e\rangle$ and $\vert f\rangle$ of the qutrit, respectively. For
simplicity, the parameters for our numerical simulation are chosen
as: $\kappa _{a}^{-1}=\kappa _{b}^{-1}=50$ $\mu$s,   $\gamma
_{g,e}^{-1}=50$ $\mu$s, $\gamma _{e,f}^{-1}=25$ $\mu$s,
$\gamma_{\phi,e}^{-1}=\gamma_{\phi,f}^{-1}=50$ $\mu$s,
$\omega_a/(2\pi )=5.5$ GHz, and $\omega_b/(2\pi )=7.0$ GHz. In the
first step, we chose $\omega _{g,e}/(2\pi )=5.5$ GHz, $\omega
_{e,f}/(2\pi )=4.7$ GHz,
$g_a^{g,e}/(2\pi)=\frac{g_a^{e,f}}{2\sqrt{2}\pi}=0.045$ GHz, and
$g_b^{g,e}/(2\pi)=\frac{g_b^{e,f}}{2\sqrt{2}\pi}=0.0005$ GHz. In the
second step, $\omega _{g,e}/(2\pi )=7.8$ GHz, $\omega _{e,f}/(2\pi
)=7.0$ GHz, $g_a^{g,e}/(2\pi)=\frac{g_a^{e,f}}{2\sqrt{2}\pi}=0.0005$
GHz, and $g_b^{g,e}/(2\pi)=\frac{g_b^{e,f}}{2\sqrt{2}\pi}=0.022$
GHz. The parameters in the third step are the same as those in the
first step. It worth noticing that the long coherence time of the
transmon qubit with $50$ $\mu$s, the high quality factor of a 1D
superconducting resonator with above $10^6$, and the tunable
coupling strength of a charge qubit and a resonator with from $200$
KHz to $43$ MHz have been realized in experiments
\cite{Srinivasan,Chang,Megrant}. For superconducting qutrits, the
typical transition frequency between two neighboring levels is from
$1$ GHz to $20$ GHz \cite{Xiang,Hoi}.

The fidelity of our c-phase gate is defined as
\begin{eqnarray}              
F=(\frac{1}{2\pi})^2\int_{0}^{2\pi}\int_{0}^{2\pi}\langle\psi_{ideal}|\rho_{f}^{c-phase}|\psi_{ideal}\rangle
d\theta_1 d\theta_2, \label{fidelity}
\end{eqnarray}
where $|\psi_{ideal}\rangle$ is the final state $\left\vert
\psi_{f}\right\rangle$ of the system composed of the resonator
qubits $r_{a}$ and $r_{b}$ after an ideal c-phase gate operation is
performed with the initial state $\left\vert \psi_{0}\right\rangle$,
which is obtained by not taking  the dissipation and dephasing into
account. $\rho_{f}^{c-phase}$ is the realistic density operator
after our c-phase gate operation on the initial state $\left\vert
\psi_{0}\right\rangle$. Our simulation shows that the fidelity of
our c-phase gate is 99.57\% within the operation time of about 38.1
ns. Taking $\theta_1=\theta_2=\frac{\pi}{4}$ as an example, the
density operators of the initial state and the final state are shown
in Fig. 3 (a) and (b), respectively.

In fact, by using the resonance operations, one can also construct
the swap gate on two resonator qubits  simply with our device by the
five steps shown in   Tab. 1.

\begin{table}
\centering \caption{The  steps for constructing the SWAP gate on
$r_a$ and $r_b$ with  all-resonance operations. }
\begin{tabular}{cccccccccc}
\hline\hline Step &  &  & Coupling &  &  & Time &  &  &
\;\;\;\;\;\;\;\; Transition \\ \hline

$\;\;$ i) &  &  & $\;\;\;\;$ $g_{a}^{g,e}$ &  &  &  $\frac{\pi
}{2g_{a}^{g,e}}$ &  &  & $ \left\vert 0\right\rangle _{a}\text{,
}\left\vert 1\right\rangle _{a}\rightarrow \left\vert g\right\rangle
_{q}\text{, }\left\vert e\right\rangle _{q}$ \\ 

$\;\;$ ii) &  & & $\;\;\;\;$ $g_{b}^{e,f}$ &  &  & $\frac{\pi
}{2g_{b}^{e,f}}$ &  &  & $ \left\vert 0\right\rangle _{b}\text{,
}\left\vert 1\right\rangle _{b}\rightarrow \left\vert e\right\rangle
_{q}\text{, }\left\vert f\right\rangle _{q}$ \\ 

$\;\;$ iii) & & & $\;\;\;\;$ $g_{b}^{g,e}$ &  &  & $\frac{3\pi
}{2g_{b}^{g,e}}$ & &  & $ \left\vert 0\right\rangle _{b}\text{,
}\left\vert 1\right\rangle _{b}\rightarrow \left\vert g\right\rangle
_{q}\text{, }\left\vert e\right\rangle _{q}$ \\ 

$\;\;$ iv) &  & & $\;\;\;\;$  $g_{b}^{e,f}$ &  &  & $\frac{\pi
}{2g_{b}^{e,f}}$ &  &  & $ \left\vert 0\right\rangle _{b}\text{,
}\left\vert 1\right\rangle _{b}\rightarrow \left\vert e\right\rangle
_{q}\text{, }\left\vert f\right\rangle _{q}$ \\ 

$\;\;$ v) &  & & $\;\;\;\;$ $g_{a}^{g,e}$ &  &  & $\frac{\pi
}{2g_{a}^{g,e}}$ &  &  & $ \left\vert 0\right\rangle _{a}\text{,
}\left\vert 1\right\rangle _{a}\rightarrow \left\vert g\right\rangle
_{q}\text{, }\left\vert e\right\rangle _{q}$ \\ \hline\hline
\end{tabular}\label{table1}
\end{table}

\begin{figure}[tpb]                 
\begin{center}
\includegraphics[width=7.2 cm,angle=0]{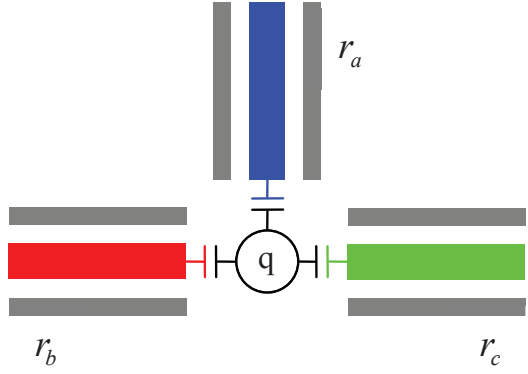} 
\end{center}
\caption{The schematic diagram for our cc-phase gate on three
resonators with all-resonance operations in circuit QED. There are
three high-quality resonators ($r_a$, $r_b$, and $r_c$) and all are
capacitively coupled to the qutrit as the same as the ones in Fig.
1.} \label{fig4}
\end{figure}

\textbf{ Cc-phase gate on superconducting resonators.}  Our cc-phase
gate is used to perform a minus phase manipulation on the three
resonator qubits only if the resonators $r_a$, $r_b$, and $r_c$ are
in the state $\vert 1\rangle_a\vert 1\rangle_b \vert 0\rangle_c$.

Our device for the cc-phase gate on the three high-quality
superconducting resonators $r_a$, $r_b$, and $r_c$ which are coupled
to the transmon qutrit $q$ is shown in Fig. 4. In the interaction
picture, the Hamiltonian of the whole system composed of the three
resonators and the qutrit is:
\begin{eqnarray}             
\begin{split}
H_{3q}  =&  g_{a}^{g,e}a\sigma _{g,e}^{+}e^{i\delta_{a}^{g,e}t}+g_{a}^{e,f}a\sigma _{e,f}^{+}e^{i\delta_{a}^{e,f}t} +g_{b}^{g,e}b\sigma _{g,e}^{+}e^{i\delta_{b}^{g,e}t}  \\
& +g_{b}^{e,f}b\sigma _{e,f}^{+}e^{i\delta_{b}^{e,f}t}
+g_{c}^{g,e}c\sigma
_{g,e}^{+}e^{i\delta_{c}^{g,e}t}+g_{c}^{e,f}c\sigma
_{e,f}^{+}e^{i\delta_{c}^{e,f}t}+h.c.. \label{hamiltonian}
\end{split}
\end{eqnarray}
Suppose that the initial state of the  system is
\begin{eqnarray}          
\begin{split} \left\vert \Psi \right\rangle_{0}=\; & (\cos\theta_1|0\rangle_a+\sin\theta_1|1\rangle_a)\otimes(\cos\theta_2|0\rangle_b+\sin\theta_2|1\rangle_b)
\otimes(\cos\theta_3|0\rangle_c+\sin\theta_3|1\rangle_c)\otimes
\left\vert g\right\rangle \!_{q}
\\=\;& (\beta_1\left\vert 0\right\rangle \!_{a}\!\left\vert
0\right\rangle \!_{b}\!\left\vert 0\right\rangle
\!_{c}+\beta_2\left\vert 0\right\rangle \!_{a}\!\left\vert
0\right\rangle \!_{b}\!\left\vert 1\right\rangle
\!_{c}+\beta_3\left\vert 0\right\rangle \!_{a}\!\left\vert
1\right\rangle \!_{b}\!\left\vert 0\right\rangle\! _{c}    +
\beta_4\left\vert 0\right\rangle \!_{a}\!\left\vert 1\right\rangle
\!_{b}\!\left\vert 1\right\rangle \!_{c} \\
&+ \beta_5\left\vert 1\right\rangle \!_{a}\!\left\vert
0\right\rangle \!_{b}\!\left\vert 0\right\rangle
\!_{c}+\beta_6\left\vert 1\right\rangle \!_{a}\!\left\vert
0\right\rangle \!_{b}\!\left\vert 1\right\rangle \!_{c}   +
\beta_7\left\vert 1\right\rangle \!_{a}\!\left\vert 1\right\rangle
\!_{b}\!\left\vert 0\right\rangle \!_{c} +\beta_8\left\vert
1\right\rangle \!_{a}\!\left\vert 1\right\rangle \!_{b}\!\left\vert
1\right\rangle \!_{c})\otimes \left\vert g\right\rangle \!_{q}.
\label{ccphase0}
\end{split}
\end{eqnarray}
Here, $\beta_1=\cos\theta_1\cos\theta_2\cos\theta_3$,
$\beta_2=\cos\theta_1\cos\theta_2\sin\theta_3$,
$\beta_3=\cos\theta_1\sin\theta_2\cos\theta_3$,
$\beta_4=\cos\theta_1\sin\theta_2\sin\theta_3$,
$\beta_5=\sin\theta_1\cos\theta_2\cos\theta_3$,
$\beta_6=\sin\theta_1\cos\theta_2\sin\theta_3$,
$\beta_7=\sin\theta_1\sin\theta_2\cos\theta_3$, and
$\beta_8=\sin\theta_1\sin\theta_2\sin\theta_3$. The cc-phase gate on
three resonator qubits can be constructed with nine resonance
operations between the qutrit and the resonators. The detailed steps
are described as follows.

First, turning off the interaction between $q$ and $r_b$ and that
between $q$ and $r_c$,  and resonating $r_a$ and $q$  with the
transition $\vert g\rangle_q \leftrightarrow\vert e\rangle_q$
($\omega_a = \omega_{g,e}$), the state of the whole system becomes
\begin{eqnarray}                   
\begin{split}
\left\vert \Psi \right\rangle_{1}    =\;& \beta_1\left\vert
0\right\rangle \!_{a}\!\left\vert 0\right\rangle \!_{b}\!\left\vert
0\right\rangle \!_{c}\!\left\vert g\right\rangle \!_{q} +
\beta_2\left\vert 0\right\rangle \!_{a}\!\left\vert 0\right\rangle
\!_{b}\!\left\vert 1\right\rangle \!_{c}\!\left\vert g\right\rangle
\!_{q}     + \beta_3\left\vert 0\right\rangle \!_{a}\!\left\vert
1\right\rangle \!_{b}\!\left\vert 0\right\rangle \!_{c}\!\left\vert
g\right\rangle \!_{q} + \beta_4\left\vert 0\right\rangle
\!_{a}\!\left\vert 1\right\rangle \!_{b}\!\left\vert
1\right\rangle \!_{c}\!\left\vert g\right\rangle \!_{q}  \\
&  - i\beta_5\left\vert 0\right\rangle \!_{a}\!\left\vert
0\right\rangle \!_{b}\!\left\vert 0\right\rangle \!_{c}\!\left\vert
e\right\rangle \!_{q}- i\beta_6\left\vert 0\right\rangle
\!_{a}\!\left\vert 0\right\rangle \!_{b}\!\left\vert 1\right\rangle
\!_{c}\!\left\vert e\right\rangle \!_{q}    -i\beta_7\left\vert
0\right\rangle \!_{a}\!\left\vert 1\right\rangle \!_{b}\!\left\vert
0\right\rangle \!_{c}\!\left\vert e\right\rangle \!_{q}-
i\beta_8\left\vert 0\right\rangle \!_{a}\!\left\vert 1\right\rangle
\!_{b}\!\left\vert 1\right\rangle \!_{c}\!\left\vert e\right\rangle
\!_{q} \label{ccphase1} \end{split}
\end{eqnarray}
after the interaction time of $t=\frac{\pi}{2g^{g,e}_a}$.

Second, turning off the interaction between $q$ and $r_a$ and that
between $q$ and $r_c$, and tuning the frequency of $r_b$ or $q$ to
make $\omega_b = \omega_{e,f}$, one can complete the resonance
manipulation on $r_b$ and $q$ with the transition $\vert e\rangle_q
\leftrightarrow\vert f\rangle_q$. The state of the whole system can
be changed into
\begin{eqnarray}                  
\begin{split}
\left\vert \Psi \right\rangle_{2}   =\;& \beta_1\left\vert
0\right\rangle \!_{a}\!\left\vert 0\right\rangle \!_{b}\!\left\vert
0\right\rangle \!_{c}\!\left\vert g\right\rangle
\!_{q}+\beta_2\left\vert 0\right\rangle \!_{a}\!\left\vert
0\right\rangle \!_{b}\!\left\vert 1\right\rangle \!_{c}\!\left\vert
g\right\rangle \!_{q}    + \beta_3\left\vert 0\right\rangle
\!_{a}\!\left\vert 1\right\rangle \!_{b}\!\left\vert 0\right\rangle
\!_{c}\!\left\vert g\right\rangle \!_{q} + \beta_4\left\vert
0\right\rangle \!_{a}\!\left\vert 1\right\rangle \!_{b}\!\left\vert
1\right\rangle \!_{c}\!\left\vert g\right\rangle \!_{q}  \\
&  - i\beta_5\left\vert 0\right\rangle \!_{a}\!\left\vert
0\right\rangle \!_{b}\!\left\vert 0\right\rangle \!_{c}\!\left\vert
e\right\rangle \!_{q} - i\beta_6\left\vert 0\right\rangle
\!_{a}\!\left\vert 0\right\rangle \!_{b}\!\left\vert 1\right\rangle
\!_{c}\!\left\vert e\right\rangle \!_{q}    - \beta_7\left\vert
0\right\rangle \!_{a}\!\left\vert 0\right\rangle \!_{b}\!\left\vert
0\right\rangle \!_{c}\!\left\vert f\right\rangle \!_{q} -
\beta_8\left\vert 0\right\rangle \!_{a}\!\left\vert 0\right\rangle
\!_{b}\!\left\vert 1\right\rangle \!_{c}\!\left\vert f\right\rangle
\!_{q}  \label{ccphase2}
\end{split}
\end{eqnarray}
after the operation time of  $t=\frac{\pi}{2g^{e,f}_2}$.

Third, repeating the same operation as the one in the first step,
the state of the whole system can be evolved into
\begin{eqnarray}                       
\begin{split}
\left\vert \Psi \right\rangle_{3}    =\;& \beta_1\left\vert
0\right\rangle \!_{a}\!\left\vert 0\right\rangle \!_{b}\!\left\vert
0\right\rangle \!_{c}\!\left\vert g\right\rangle \!_{q}  +
\beta_2\left\vert 0\right\rangle \!_{a}\!\left\vert 0\right\rangle
\!_{b}\!\left\vert 1\right\rangle \!_{c}\!\left\vert g\right\rangle
\!_{q}  + \beta_3\left\vert 0\right\rangle \!_{a}\!\left\vert
1\right\rangle \!_{b}\!\left\vert 0\right\rangle \!_{c}\!\left\vert
g\right\rangle \!_{q} + \beta_4\left\vert 0\right\rangle
\!_{a}\!\left\vert 1\right\rangle \!_{b}\!\left\vert
1\right\rangle \!_{c}\!\left\vert g\right\rangle \!_{q}  \\
&  - \beta_5\left\vert 1\right\rangle \!_{a}\!\left\vert
0\right\rangle \!_{b}\!\left\vert 0\right\rangle \!_{c}\!\left\vert
g\right\rangle \!_{q} - \beta_6\left\vert 1\right\rangle
\!_{a}\!\left\vert 0\right\rangle \!_{b}\!\left\vert 1\right\rangle
\!_{c}\!\left\vert g\right\rangle \!_{q}    - \beta_7\left\vert
0\right\rangle \!_{a}\!\left\vert 0\right\rangle \!_{b}\!\left\vert
0\right\rangle \!_{c}\!\left\vert f\right\rangle \!_{q} -
\beta_8\left\vert 0\right\rangle \!_{a}\!\left\vert 0\right\rangle
\!_{b}\!\left\vert 1\right\rangle \!_{c}\!\left\vert f\right\rangle
\!_{q}.  \label{ccphase3}
\end{split}
\end{eqnarray}

Fourth, turning off the interaction between $q$ and $r_b$ and that
between $q$  and  $r_c$, and resonating $r_a$ and $q$ with the
transition $\vert e\rangle_q \leftrightarrow\vert f\rangle_q$, the
state of the whole system evolves from $\left\vert \Psi
\right\rangle_{3} $  into
\begin{eqnarray}                  
\begin{split}
\left\vert \Psi \right\rangle_{4}    =\;& \beta_1\left\vert
0\right\rangle \!_{a}\!\left\vert 0\right\rangle \!_{b}\!\left\vert
0\right\rangle \!_{c}\!\left\vert g\right\rangle \!_{q} +
\beta_2\left\vert 0\right\rangle \!_{a}\!\left\vert 0\right\rangle
\!_{b}\!\left\vert 1\right\rangle \!_{c}\!\left\vert g\right\rangle
\!_{q}     + \beta_3\left\vert 0\right\rangle \!_{a}\!\left\vert
1\right\rangle \!_{b}\!\left\vert 0\right\rangle \!_{c}\!\left\vert
g\right\rangle \!_{q} + \beta_4\left\vert 0\right\rangle
\!_{a}\!\left\vert 1\right\rangle \!_{b}\!\left\vert
1\right\rangle \!_{c}\!\left\vert g\right\rangle \!_{q}  \\
&  - \beta_5\left\vert 1\right\rangle \!_{a}\!\left\vert
0\right\rangle \!_{b}\!\left\vert 0\right\rangle \!_{c}\!\left\vert
g\right\rangle \!_{q} - \beta_6\left\vert 1\right\rangle
\!_{a}\!\left\vert 0\right\rangle \!_{b}\!\left\vert 1\right\rangle
\!_{c}\!\left\vert g\right\rangle \!_{q}    + \,i \beta_7\left\vert
1\right\rangle \!_{a}\!\left\vert 0\right\rangle \!_{b}\!\left\vert
0\right\rangle \!_{c}\!\left\vert e\right\rangle \!_{q} + i
\beta_8\left\vert 1\right\rangle \!_{a}\!\left\vert 0\right\rangle
\!_{b}\!\left\vert 1\right\rangle \!_{c}\!\left\vert e\right\rangle
\!_{q}  \label{ccphase4}
\end{split}
\end{eqnarray}
after the operation time of $t=\frac{\pi}{2g^{e,f}_a}$.

Fifth, turning off the interaction between $q$ and $r_a$ and that
between $q$ and $r_b$, and resonating $r_c$ and $q$ with the
transition $\vert e\rangle_q \leftrightarrow\vert f\rangle_q$
($\omega_c = \omega_{e,f}$), the state becomes
\begin{eqnarray}                       
\begin{split}
\left\vert \Psi \right\rangle_{5}    =\;& \beta_1\left\vert
0\right\rangle \!_{a}\!\left\vert 0\right\rangle \!_{b}\!\left\vert
0\right\rangle \!_{c}\!\left\vert g\right\rangle \!_{q} +
\beta_2\left\vert 0\right\rangle \!_{a}\!\left\vert 0\right\rangle
\!_{b}\!\left\vert 1\right\rangle \!_{c}\!\left\vert g\right\rangle
\!_{q}    + \beta_3\left\vert 0\right\rangle \!_{a}\!\left\vert
1\right\rangle \!_{b}\!\left\vert 0\right\rangle \!_{c}\!\left\vert
g\right\rangle \!_{q} + \beta_4\left\vert 0\right\rangle
\!_{a}\!\left\vert 1\right\rangle \!_{b}\!\left\vert
1\right\rangle \!_{c}\!\left\vert g\right\rangle \!_{q}  \\
&   - \beta_5\left\vert 1\right\rangle \!_{a}\!\left\vert
0\right\rangle \!_{b}\!\left\vert 0\right\rangle \!_{c}\!\left\vert
g\right\rangle \!_{q} - \beta_6\left\vert 1\right\rangle
\!_{a}\!\left\vert 0\right\rangle \!_{b}\!\left\vert 1\right\rangle
\!_{c}\!\left\vert g\right\rangle \!_{q}   +\,i\beta_7\left\vert
1\right\rangle \!_{a}\!\left\vert 0\right\rangle \!_{b}\!\left\vert
0\right\rangle \!_{c}\!\left\vert e\right\rangle \!_{q} -
i\beta_8\left\vert 1\right\rangle \!_{a}\!\left\vert 0\right\rangle
\!_{b}\!\left\vert 1\right\rangle \!_{c}\!\left\vert e\right\rangle
\!_{q}  \label{ccphase5}
\end{split}
\end{eqnarray}
after the operation time of $g^{e,f}_c t=\pi$.

\begin{figure}[tpb!]
\begin{center}
\includegraphics[width=13.6 cm,angle=0]{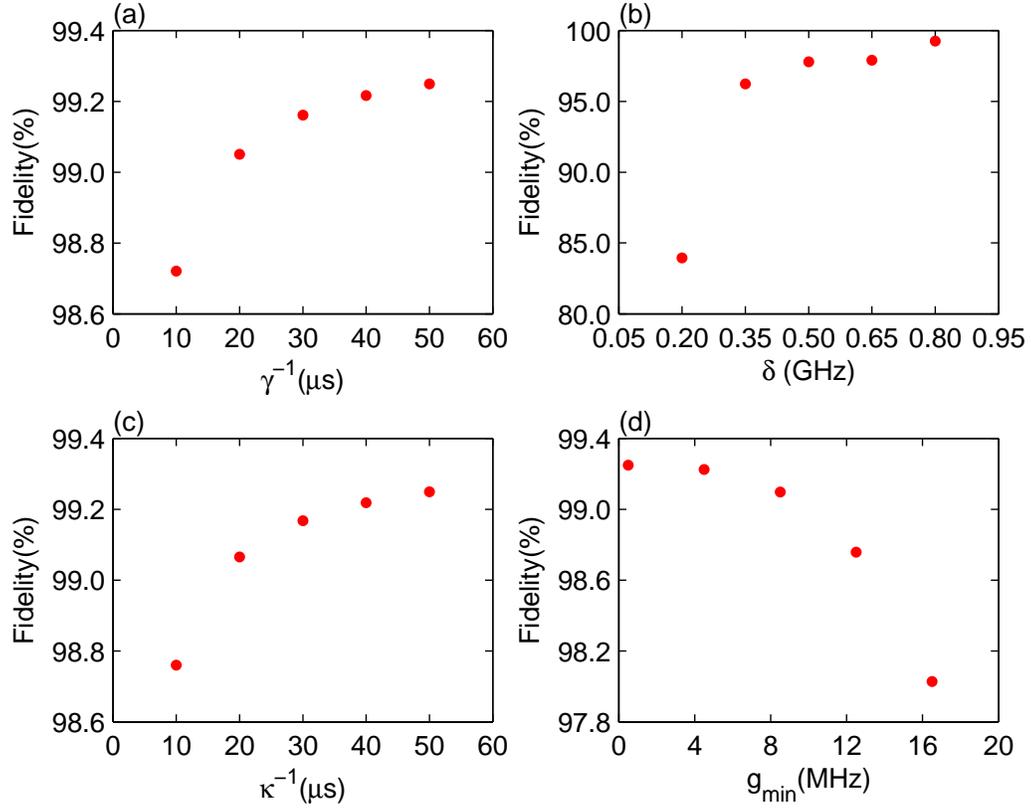}
\end{center}
\caption{ The density operator $\rho_f^{cc-phase}$ of the final
state $\left\vert \Psi_{f}\right\rangle$ of the system composed of
the three resonators and the qutrit for constructing our cc-phase
gate on three microwave-photon resonators.} \label{fig5}
\end{figure}

Sixth, repeating the fourth step, the state of the whole system
becomes
\begin{eqnarray}                 
\begin{split}
\left\vert \Psi \right\rangle_{6}   =\;& \beta_1\left\vert
0\right\rangle \!_{a}\!\left\vert 0\right\rangle \!_{b}\!\left\vert
0\right\rangle \!_{c}\!\left\vert g\right\rangle \!_{q} +
\beta_2\left\vert 0\right\rangle \!_{a}\!\left\vert 0\right\rangle
\!_{b}\!\left\vert 1\right\rangle \!_{c}\!\left\vert g\right\rangle
\!_{q}   + \beta_3\left\vert 0\right\rangle \!_{a}\!\left\vert
1\right\rangle \!_{b}\!\left\vert 0\right\rangle \!_{c}\!\left\vert
g\right\rangle \!_{q} + \beta_4\left\vert 0\right\rangle
\!_{a}\!\left\vert 1\right\rangle \!_{b}\!\left\vert
1\right\rangle \!_{c}\!\left\vert g\right\rangle \!_{q} \\
&   -\, \beta_5\left\vert 1\right\rangle \!_{a}\!\left\vert
0\right\rangle \!_{b}\!\left\vert 0\right\rangle \!_{c}\!\left\vert
g\right\rangle \!_{q} - \beta_6\left\vert 1\right\rangle
\!_{a}\!\left\vert 0\right\rangle \!_{b}\!\left\vert 1\right\rangle
\!_{c}\!\left\vert g\right\rangle \!_{q}    + \beta_7\left\vert
0\right\rangle \!_{a}\!\left\vert 0\right\rangle \!_{b}\!\left\vert
0\right\rangle \!_{c}\!\left\vert f\right\rangle \!_{q} -
\beta_8\left\vert 0\right\rangle \!_{a}\!\left\vert 0\right\rangle
\!_{b}\!\left\vert 1\right\rangle \!_{c}\!\left\vert f\right\rangle
\!_{q}. \label{ccphase6}
\end{split}
\end{eqnarray}

Seventh, taking the same manipulation as the one in the first step,
the system is in the state
\begin{eqnarray}                
\begin{split}
\left\vert \Psi \right\rangle_{7}   =\;& \beta_1\left\vert
0\right\rangle \!_{a}\!\left\vert 0\right\rangle \!_{b}\!\left\vert
0\right\rangle \!_{c}\!\left\vert g\right\rangle \!_{q} +
\beta_2\left\vert 0\right\rangle \!_{a}\!\left\vert 0\right\rangle
\!_{b}\!\left\vert 1\right\rangle \!_{c}\!\left\vert g\right\rangle
\!_{q}   + \beta_3\left\vert 0\right\rangle \!_{a}\!\left\vert
1\right\rangle \!_{b}\!\left\vert 0\right\rangle \!_{c}\!\left\vert
g\right\rangle \!_{q}+ \beta_4\left\vert 0\right\rangle
\!_{a}\!\left\vert 1\right\rangle \!_{b}\!\left\vert
1\right\rangle \!_{c}\!\left\vert g\right\rangle \!_{q}  \\
&  +  i\beta_5\left\vert 0\right\rangle \!_{a}\!\left\vert
0\right\rangle \!_{b}\!\left\vert 0\right\rangle \!_{c}\!\left\vert
e\right\rangle \!_{q} +i\beta_6\left\vert 0\right\rangle
\!_{a}\!\left\vert 0\right\rangle \!_{b}\!\left\vert 1\right\rangle
\!_{c}\!\left\vert e\right\rangle \!_{q}    + \beta_7\left\vert
0\right\rangle \!_{a}\!\left\vert 0\right\rangle \!_{b}\!\left\vert
0\right\rangle \!_{c}\!\left\vert f\right\rangle \!_{q} -
\beta_8\left\vert 0\right\rangle \!_{a}\!\left\vert 0\right\rangle
\!_{b}\!\left\vert 1\right\rangle \!_{c}\!\left\vert f\right\rangle
\!_{q}.  \label{ccphase7}
\end{split}
\end{eqnarray}

Eighth, repeating  the second step, the state of the system evolves
from $\left\vert \Psi \right\rangle_{7}$ into
\begin{eqnarray}                   
\begin{split}
\left\vert \Psi \right\rangle \!_{8}   =\;& \beta_1\left\vert
0\right\rangle \!_{a}\!\left\vert 0\right\rangle \!_{b}\!\left\vert
0\right\rangle \!_{c}\!\left\vert g\right\rangle \!_{q} +
\beta_2\left\vert 0\right\rangle \!_{a}\!\left\vert 0\right\rangle
\!_{b}\!\left\vert 1\right\rangle \!_{c}\!\left\vert g\right\rangle
\!_{q}    + \beta_3\left\vert 0\right\rangle \!_{a}\!\left\vert
1\right\rangle \!_{b}\!\left\vert 0\right\rangle \!_{c}\!\left\vert
g\right\rangle \!_{q} + \beta_4\left\vert 0\right\rangle
\!_{a}\!\left\vert 1\right\rangle \!_{b}\!\left\vert
1\right\rangle \!_{c}\!\left\vert g\right\rangle \!_{q}  \\
&  +  i\beta_5\left\vert 0\right\rangle \!_{a}\!\left\vert
0\right\rangle \!_{b}\!\left\vert 0\right\rangle \!_{c}\!\left\vert
e\right\rangle \!_{q} + i\beta_6\left\vert 0\right\rangle
\!_{a}\!\left\vert 0\right\rangle \!_{b}\!\left\vert 1\right\rangle
\!_{c}\!\left\vert e\right\rangle \!_{q}     -  i\beta_7\left\vert
0\right\rangle \!_{a}\!\left\vert 1\right\rangle \!_{b}\!\left\vert
0\right\rangle \!_{c}\!\left\vert e\right\rangle \!_{q} +
i\beta_8\left\vert 0\right\rangle \!_{a}\!\left\vert 1\right\rangle
\!_{b}\!\left\vert 1\right\rangle \!_{c}\!\left\vert e\right\rangle
\!_{q}. \label{ccphase8}
\end{split}
\end{eqnarray}

Ninth, repeating the first step, one can get the final state of the
whole system as follows
\begin{eqnarray}                  
\begin{split}
\left\vert \Psi \right\rangle \!_{f}  =\;& (\beta_1\left\vert
0\right\rangle \!_{a}\!\left\vert 0\right\rangle \!_{b}\!\left\vert
0\right\rangle \!_{c} + \beta_2\left\vert 0\right\rangle
\!_{a}\!\left\vert 0\right\rangle \!_{b}\!\left\vert 1\right\rangle
\!_{c} + \beta_3\left\vert 0\right\rangle \!_{a}\!\left\vert
1\right\rangle \!_{b}\!\left\vert 0\right\rangle \!_{c}    + \beta_4
\left\vert 0\right\rangle \!_{a}\!\left\vert 1\right\rangle
\!_{b}\!\left\vert 1\right\rangle \!_{c} \\
& + \beta_5\left\vert 1\right\rangle \!_{a}\!\left\vert
0\right\rangle \!_{b}\!\left\vert 0\right\rangle \!_{c} +
\beta_6\left\vert 1\right\rangle \!_{a}\!\left\vert 0\right\rangle
\!_{b}\!\left\vert 1\right\rangle \!_{c}    - \beta_7\left\vert
1\right\rangle \!_{a}\!\left\vert 1\right\rangle \!_{b}\!\left\vert
0\right\rangle \!_{c} + \beta_8\left\vert 1\right\rangle
\!_{a}\!\left\vert 1\right\rangle \!_{b}\!\left\vert 1\right\rangle
\!_{c})\otimes \left\vert g\right\rangle \!_{q}. \label{ccphase9}
\end{split}
\end{eqnarray}
This is just the result of our cc-phase gate on the three
microwave-photon  resonators.

The evolution of the system composed of three resonators coupled to
the transmon  qutrit can be described by the master equation
\begin{eqnarray}              
\begin{split}
\frac{d\rho }{dt}  =&  -i[H_{3q},\rho ]+\kappa_a D[a]\rho + \kappa_b
D[b]\rho + \kappa_c D[c]\rho
+\gamma_{g,e}D[\sigma_{g,e}^{-}] \rho +\gamma_{e,f}D[\sigma_{e,f}^{-}]\rho\\
&   +
\gamma_{\phi,e}(\sigma_{ee}\rho\sigma_{ee}-\sigma_{ee}\rho/2-\rho\sigma_{ee}/2)
+
\gamma_{\phi,f}(\sigma_{ff}\rho\sigma_{ff}-\sigma_{ff}\rho/2-\rho\sigma_{ff}/2).
\label{masterequation}
\end{split}
\end{eqnarray}
Here $\kappa_c$  is the decay rate of the resonator  $r_c$. In our
simulation for the fidelity of our cc-phase gate, the parameters of
the system are chosen as: $\omega_a/(2\pi)=5.5$ GHz,
$\omega_b/(2\pi)=7.0$ GHz, $\omega_c/(2\pi)=8.0$ GHz,  $\kappa
_{a}^{-1}=\kappa _{b}^{-1}=\kappa _{c}^{-1}=50$ $\mu$s, and
$\omega_{g,e}/(2\pi)-\omega_{e,f}/(2\pi)=800$ MHz. The energy
relaxation rates and the dephasing rates of the transmon qutrit are
chosen the same as those in the construction of our c-phase gate.
The details for the parameters chosen in each step for the
simulation of our cc-phase gate are shown in Table. 2.

Let us define the fidelity of our cc-phase gate as
\begin{eqnarray}              
F=(\frac{1}{2\pi})^3\int_{0}^{2\pi}\int_{0}^{2\pi}\int_{0}^{2\pi}\langle\psi_{ideal}|\rho_{f}^{cc-phase}|\psi_{ideal}\rangle
d\theta_1 d\theta_2 d\theta_3, \label{fidelity}
\end{eqnarray}
where $|\psi_{ideal}\rangle$ is the final state $\left\vert
\Psi\right\rangle_{f}$ of the system composed of three resonator
qubits $r_{a}$, $r_{b}$, and $r_{c}$ after an ideal cc-phase gate
operation when the initial state of the system is $\left\vert
\Psi\right\rangle_{0}$, without considering the dissipation and the
dephasing. $\rho_{f}^{cc-phase}$ is the realistic density operator
after our cc-phase gate operation on the initial state $\left\vert
\Psi\right\rangle_{0}$. We numerically simulate the fidelity of our
cc-phase gate, by taking the dissipation and the dephasing into
account. The fidelity of our cc-phase gate is 99.25\% within the
operation time of about 73.3 ns.

In a realistic experiment, the energy relaxation rate $\gamma$ and
the anharmonicity $\delta=\omega_{g,e}-\omega_{e,f}$ of the qutrit,
the decay rate $\kappa$ of the resonator, and the minimum value of
tunable coupling strength $g_{min}$  influence the fidelity of our
cc-phase gate. Their effects are shown in Fig. 5 (a)-(d) in which we
simulate the fidelity of the gate by varying a single parameter and
fixing  the other parameters. In Fig. 5 (b), although the fidelity
of the cc-phase gate is reduced obviously when the anharmonicity of
the qutrit becomes small, it can in principle be improved by taking
a smaller coupling strength for the resonance operation.

\bigskip

{\large \textbf{Discussion}}

The number-state-dependent interaction between a superconducting
qubit and resonator qudits is an important nonlinear effect which
has been used to construct the quantum entangled states and quantum
logic gates on resonator qudits in the previous works
\cite{FWStrauch,frederick,Wu,Hua}. This effect is a useful
second-order coupling between the qubit and the resonator in the
dispersive regime, which indicates a slow operation of the
state-dependent selective rotation on the qubit with a drive field.
In contrary, our gates are achieved by using the quantum resonance
operation only, which is not the high-order coupling item of the
qubit and the resonator, and has been realized for generating the
Fock states in a superconducting resonator with a high fidelity
\cite{Hofheinz}.  All-resonance-based quantum operations make our
universal quantum gates on microwave-photon resonators have a
shorter operation time, compared with those in previous works
\cite{Hua}.  Moreover, our gates have a higher fidelity than those
in the latter if we take the decoherence of the qubit and the decay
of the resonators into account. Although there are nine steps in
constructing our cc-phase gate on three resonators, compared with
the three steps in constructing our c-phase gate, the total period
of the resonance operations in our cc-phase gate is not much longer
than the one in our c-phase gate.

In our simulations, the quantum errors from the preparation of the
initial states of Eqs.(\ref{cphase0}) and  (\ref{ccphase0}) are not
considered. Single-qubit operations on a qubit \cite{Barends} have
been realized with the error  smaller than $10^{-4}$ and it can be
depressed to   much small \cite{Motzoi}. That is to say, the error
from single-qubit operations  has only a negligible influence on the
results of the fidelities of our fast universal quantum gates. There
are several methods which can help us to turn on and off the
resonance interaction between a superconducting qutrit and a
resonator, such as tuning the frequency of the qutrit, tuning the
frequency of the resonator, or tuning their coupling strength. In
experiment, a tunable coupling superconducting device has been
realized \cite{Srinivasan,Allman}. The coupling strength between a
phase qubit and a lumped element resonator \cite{Allman} can be
tuned from $0$ MHz to $100$ MHz. The coupling strength between a
charge qubit and a resonator \cite{Srinivasan} can be tuned from
$200$ KHz to $43$ MHz. Tuning the frequency of a high quality
resonator has also been realized \cite{Sandberg}. The frequency of a
1D superconducting resonator with the quality of $10^4$ can be tuned
with a range of $740$ MHz. The frequency of a transmon qubit
\cite{Schreier}  can be tuned in a range of about $2.5$ GHz. In the
system composed of several resonators coupled to a superconducting
qutrit, by tuning the frequency of the qutrit only to complete the
resonance operation between the qutrit and the resonators with a
high fidelity, one should take small coupling strengths between
them, which leads  to a long-time operation \cite{Su}. By using the
tunable resonator or tunable coupling strength only to turn on and
off the interaction, the fast high-fidelity resonance operation
requires a much larger tunable range. Here, we tune the frequency of
the qutrit and the coupling strengths between the qutrit and each
resonator to turn on and off their resonance interactions to achieve
our fast universal gates. The coupling strengths are chosen much
smaller than the anharmonicity of the transmon qutrit, which helps
us to treat the qutrit as a qubit \cite{Schreier} during the
resonance operations without  considering the effect from the third
excited energy level of the qutrit. To implement our gates in
experiment with a high fidelity, one should also apply a magnetic
flux with fast tunability. On one hand, it can tune the frequency of
the qutrit instantaneously to get the high-fidelity resonance
operation \cite{Houck}. On the other hand, it can help us to get a
fast tunable coupling strength between the qutrit and the resonator
\cite{Allman,Srinivasan}.

In summary, we have proposed two schemes for the  construction of
universal quantum gates on resonator qubits in the processor
composed of multiple  high-quality microwave-photon resonators
coupled to a transmon  qutrit, including the c-phase and cc-phase
gates. Different from the ones in the previous works based on the
dispersive coupling effect of the number-state-dependent interaction
between a superconducting qubit and the resonator qubits \cite{Hua},
our gates are achieved by all-resonance quantum operations and they
have the advantages of higher fidelities and shorter operation
times.  With the optimal feasible parameters, our numerical
simulations show that the fidelity of our c-phase gate approaches
99.57\% within the operation time of   38.1 ns and that of our
cc-phase gate is 99.25\%  within  73.3 ns,  not resorting to  drive
fields.

\begin{table}
\centering \caption{The parameters for constructing the cc-phase
gate on $r_a$, $r_b$, and $r_c$. }
\begin{tabular}{lllllllllllll}
\hline\hline \multirow{2}{*}{Step}  &  & $\omega _{g,e}/(2\pi )$ &
&  $g_{a}^{g,e}/(2\pi )$
                       &  &  $g_{b}^{g,e}(2\pi )$      &  &  $g_{c}^{g,e}(2\pi )$ \\
                       &  & \;\;{\footnotesize (GHZ)}  &  &  \;\;{\footnotesize (MHZ)}
                       &  &  \;\;{\footnotesize (MHZ)} &  & \;\;{\footnotesize (MHZ)}
\\\hline
\;\;i)    &  &\;\; $5.5$ &  &\;\;  $45$ &  &\;\; $0.5$ &  &\;\;
$0.5$ \\ 
\;\;ii)   &  &\;\; $7.8$ &  &\;\; $0.5$ &  &\;\;
$28$  &  &\;\; $0.5$ \\ 
\;\;iii)  &  &\;\; $5.5$ &  &\;\;
$27$  &  &\;\; $0.5$ &  &\;\; $0.5$ \\ 
\;\;iv)   &  &\;\;
$6.3$ &  &\;\; $24$  &  &\;\; $0.5$ &  &\;\; $0.5$ \\ 
\;\;v)
&  &\;\; $8.8$ &  &\;\; $0.5$ &  &\;\; $0.5$ &  &\;\; $20$ \\ 
\;\;vi)   &  &\;\; $6.3$ &  &\;\; $29$  &  &\;\; $0.5$ &  &\;\;
$0.5$ \\ 
\;\;vii)  &  &\;\; $5.5$ &  &\;\; $27$  &  &\;\;
$0.5$ &  &\;\; $0.5$ \\ 
\;\;viii) &  &\;\; $7.8$ &  &\;\;
$0.5$ &  &\;\; $28$  &  &\;\; $0.5$ \\ 
\;\;ix)   &  &\;\;
$5.5$ &  &\;\; $45$  &  &\;\; $0.5$ &  &\;\; $0.5$ \\ \hline\hline
\end{tabular}\label{table2}
\end{table}

\begin{figure}[tpb!]                 
\begin{center}
\includegraphics[width=10.0 cm,angle=0]{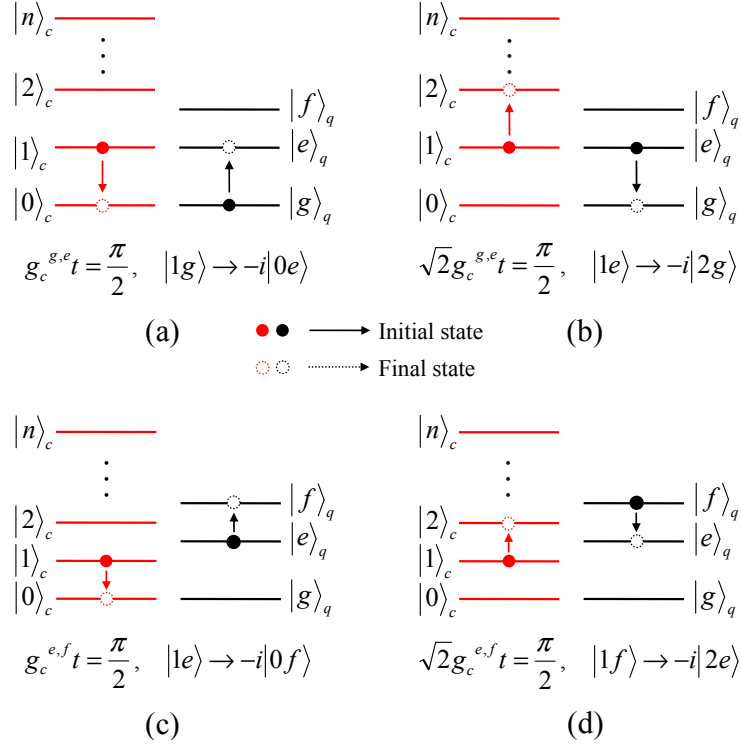} 
\end{center}
\caption{ Schematic diagram for the resonance processes between a
single-model cavity field and a qutrit. $|n\rangle_c$ is the Fock
state of the cavity. $t$ is the operation time of the resonance
processing. } \label{fig6}
\end{figure}

\bigskip\bigskip

{\large \textbf{Methods}}

\textbf{Quantum resonance operation.} Quantum resonance operation is
the key element for the construction of  our  all-resonance-based
universal quantum gates on microwave-photon resonators.  In a system
composed of a two-energy-level qubit coupled to a cavity, the
Hamiltonian of the system is (in the interaction picture)
\cite{book2}
\begin{eqnarray}             
H_{I} = g(a^{+}\sigma_{-}e^{-i\Delta t}+a\sigma_{+}e^{i\Delta t}).
\label{JC}
\end{eqnarray}
Here $\Delta=\omega_c-\omega_q$ and $\omega_c$ is the frequency of
the cavity. The Hamiltonian $H_{I}$ describes the state transfer
between the  qubit  and the cavity.  In the system, the unitary
time-evolution operation is given by $U(t)=e^{-iH_{I}t}$, which can
be expanded at the exact resonances between the qubit and the cavity
($\Delta=0$) as \cite{book2}
\begin{eqnarray}             
\begin{split}
U(t)  =\;& cos(gt\sqrt{a^{+}a+1})|e\rangle\langle e| + cos(gt\sqrt{a^{+}a})|g\rangle \langle g|   \\
&  -
i\frac{sin(gt\sqrt{a^{+}a+1})}{\sqrt{a^{+}a+1}}a|e\rangle\langle g|
-
ia^{+}\frac{sin(gt\sqrt{a^{+}a+1})}{\sqrt{a^{+}a+1}}|g\rangle\langle
e|. \label{evolution}
\end{split}
\end{eqnarray}

In our work, the resonance interactions take place between a
three-energy-level  qutrit and a single-model cavity field. To keep
the resonance operation between the cavity and the qutrit with the
wanted transition $|g\rangle \leftrightarrow |e\rangle$ or
$|e\rangle \leftrightarrow |f\rangle$,  one should take a small
coupling strength between the qutrit and the cavity, compared with
the anharmonicity of the qutrit, to avoid the off-resonance
interaction between the cavity and the qutrit with the unwanted
transition. The details of the state evolution of the system
composed of a qutrit and a cavity are described in Fig. 6 (in which
we give all the resonance processes used in this work only). In the
quantum resonance operation between the cavity and the qutrit in the
transmission between the energy levels $\vert g\rangle$ and $\vert
e\rangle$, the evolution $\vert 1g\rangle\rightarrow -i\vert
0e\rangle$ ($\vert 0e\rangle\rightarrow -i\vert 1g\rangle$) is
completed with $g_c^{g,e}t=\pi/2$, shown in Fig. 6 $(a)$. With
$\sqrt{2}g_c^{g,e}t=\pi/2$, the evolution $\vert
1e\rangle\rightarrow -i\vert 2g\rangle$ ($\vert 2g\rangle\rightarrow
-i\vert 1e\rangle$) can be achieved, shown in Fig. 6 $(b)$. In the
quantum resonance operation between the cavity and the qutrit in the
transmission between the energy levels $\vert e\rangle$ and $\vert
f\rangle$, the evolution $\vert 1e\rangle\rightarrow -i\vert
0f\rangle$ ($\vert 0f\rangle\rightarrow -i\vert 1e\rangle$) is
completed with $g_c^{e,f}t=\pi/2$, shown in Fig. 6 $(c)$. With
$\sqrt{2}g_c^{e,f}t=\pi/2$, the evolution $\vert
1f\rangle\rightarrow -i\vert 2e\rangle$ ($\vert 2e\rangle\rightarrow
-i\vert 1f\rangle$) can be achieved, shown in Fig. 6 $(d)$.


\bigskip
\bigskip

{\large \textbf{Acknowledgments}}

  This work is supported by the National Natural Science Foundation of
China under Grant Nos. 11174039 and 11474026, and NECT-11-0031.

\bigskip

{\large \textbf{Authors contributions}}

M. H. and M.J.  completed the calculation and prepared the figures.
M. H. and F.G. wrote the main manuscript text. F.G. supervised the
whole project. All authors reviewed the manuscript.

\bigskip

{\large \textbf{Additional information}}

 Competing financial interests: The authors declare no competing
financial interests.

\end{document}